\documentclass{article}

\usepackage[preprint]{neurips_2022_ml4ps}

\usepackage[utf8]{inputenc} 
\usepackage[T1]{fontenc}    
\usepackage{url}            
\usepackage{booktabs}       
\usepackage{amsfonts}       
\usepackage{nicefrac}       
\usepackage{microtype}      
\usepackage{xcolor}         
\usepackage{mathtools}
\usepackage{physics,amsmath}
\usepackage{multirow}
\usepackage{ctable}
\usepackage{multicol}
\usepackage{bm}
\usepackage{aas_macros}

\title{Semi-Supervised Domain Adaptation for Cross-Survey Galaxy Morphology Classification and Anomaly Detection}

\author{%
  Aleksandra \'Ciprijanovi\'c \\
  Fermi National Accelerator Laboratory\\
   Kirk Rd and Pine St\\
 Batavia, IL 60510 \\
  \texttt{aleksand@fnal.gov} \\
   \And
   Ashia Lewis \\
   Fermi National Accelerator Laboratory\\
   Kirk Rd and Pine St\\
   Batavia, IL 60510 \\
  \texttt{atlewis@fnal.gov} \\
   \AND
   Kevin Pedro \\
  Fermi National Accelerator Laboratory\\
   Kirk Rd and Pine St\\
 Batavia, IL 60510 \\
   \texttt{pedrok@fnal.gov} \\
   \And
   Sandeep Madireddy \\
   Mathematics and Computer Science Division,\\
   Argonne National Laboratory \\
  Lemont, IL 60439, USA \\
 \texttt{smadireddy@anl.gov} \\
   \AND
   Brian Nord \\
  Fermi National Accelerator Laboratory;\\
 Kavli Institute for Cosmological Physics \&\\ 
Department of Astronomy and Astrophysics,\\
University of Chicago; \\
Laboratory for Nuclear Physics, MIT \\
   \texttt{nord@fnal.gov} \\
   \And
      Gabriel N. Perdue \\
  Fermi National Accelerator Laboratory\\
   Kirk Rd and Pine St\\
 Batavia, IL 60510 \\
   \texttt{perdue@fnal.gov} \\
   \AND
      Stefan M. Wild \\
 Mathematics and Computer Science Division,\\
   Argonne National Laboratory \\
  Lemont, IL 60439, USA \\
 \texttt{wild@anl.gov} \\
}

\begin{document}

\maketitle

\begin{abstract}
In the era of big astronomical surveys, our ability to leverage artificial intelligence algorithms simultaneously for multiple datasets will open new avenues for scientific discovery. Unfortunately, simply training a deep neural network on images from one data domain often leads to very poor performance on any other dataset. Here we develop a Universal Domain Adaptation method \textit{DeepAstroUDA}, capable of performing semi-supervised domain alignment that can be applied to datasets with different types of class overlap. Extra classes can be present in any of the two datasets, and the method can even be used in the presence of unknown classes. For the first time, we demonstrate the successful use of domain adaptation on two very different observational datasets (from SDSS and DECaLS). We show that our method is capable of bridging the gap between two astronomical surveys, and also performs well for anomaly detection and clustering of unknown data in the unlabeled dataset. We apply our model to two examples of galaxy morphology classification tasks with anomaly detection: 1) classifying spiral and elliptical galaxies with detection of merging galaxies (three classes including one unknown anomaly class); 2) a more granular problem where the classes describe more detailed morphological properties of galaxies, with the detection of gravitational lenses (ten classes including one unknown anomaly class).
\end{abstract}

\section{Introduction}
With big datasets from astronomical surveys like the Dark Energy Survey [DES; 5] or the Vera Rubin Legacy Survey of space and time [LSST; 9], development of artificial intelligence (AI) algorithms capable of combining knowledge from different telescopes will open doors to many new insights. Unfortunately, standard deep learning algorithms are not well suited for work on multiple datasets. Domain adaptation (DA) research includes the development of methods designed to enable work on multiple datasets at the same time, by allowing the model to learn and use only features present in both datasets [4, 18, 20], thus aligning two latent data distributions. In astronomy, DA has first been used in combination with active learning in [16], for Supernova Ia classification and identification of Mars landforms. Then, in [2], authors use DA for identifying merging galaxies in simulated Sloan Digital Sky Survey [SDSS; 21] mock images and and real SDSS data. In [3], authors use DA for galaxy morphology classification in simulated mock LSST data, and as a tool to mitigate possible catastrophic perturbation-driven errors. Unfortunately, even when two datasets are quite similar, like simulated mock images mimicking some telescope and real observations from the same telescope, DA methods can be quite hard to fine tune. Because of this, the development of DA methods that can be used on multiple observational datasets exhibiting even larger differences has not previously been attempted. Furthermore, when working with real astronomical data, researchers will not always be able to work with fully curated datasets; there may be partial overlap of classes, unknown classes or anomalies in one or both domains, etc. Most standard DA methods try to align the entire data distributions, so the presence of any kind of non-overlapping classes will not allow those DA methods to be applied successfully. 

In this work we develop a semi-supervised universal DA method \textit{DeepAstroUDA} that can handle the presence of non-overlapping classes in any of the two data domains and can even detect and cluster unknown classes. As with all DA methods, it requires two datasets: the source domain, which contains labeled images, and the target domain, which can be unlabeled as the labels are not used during model training. Our aim is the development of a universal DA method that can be applied to a plethora of astronomical tasks and can successfully perform DA on both simulated and observational data. We apply our method to two observational datasets (available via the Galaxy Zoo project [12, 11, 19]) and show that it is capable of successfully bridging even two substantially different datasets and even discovering new unknown classes in the unlabeled target domain. We focus on galaxy morphology classification with anomaly detection: 1) classification of spiral and elliptical galaxies, with the discovery of merging galaxies, and 2) more granular galaxy morphology classification with sub-classes that more closely describe galaxy shapes (based on attributes such as ellipticity, bulge prominence, presence of a bar in spiral galaxies etc.) and detection of gravitationally lensed galaxies.

\section{Methods}\label{sec:methods}

In this work, we introduce a semi-supervised universal DA method \textit{DeepAstroUDA}. Domain alignment and clustering of similar objects into classes is performed via two loss functions: adaptive clustering and entropy separation\footnote{Code is publicly available at: \url{https://github.com/deepskies/DeepAstroUDA}}. 

\textbf{Adaptive Clustering (AC) Loss}: The main idea of this type of semi-supervised clustering [10]is to group unlabeled target domain samples into clusters by computing pairwise similarities among their extracted features, then force the class labels predicted by the classifier for samples with large pairwise feature similarities to be consistent.   
For a pair of unlabeled target samples $x_1$ and $x_2$ we predict a pairwise similarity label by using the output prediction vectors $\vectorbold p_1$ and $\vectorbold p_2$ and rank ordering their feature elements [7]. We then require the top $k$ elements ($k=3$ for 3-class problem and $k=7$ for 10-class problem, for computational efficiency) to decide that the paired samples belong to the same class, which is denoted by similarity label $s_{12} = 1$; otherwise, $s_{12} = 0$. For the labeled source domain images, we use their class labels to generate similarity labels.
We also calculate a similarity score between samples as $\vectorbold p^\top_1 \vectorbold p_2$. Finally, we can write the AC loss as a binary cross-entropy loss, where similarity labels are used as ground truth labels:

\begin{equation}
{\cal L}_\mathrm{AC} = -\sum_{i \in B}\sum_{j \in b_t} s_{ij}\mathrm{log}(\vectorbold p^\top_i \vectorbold p_j) + (1-s_{ij})\mathrm{log}(1-\vectorbold p^\top_i \vectorbold p_j),
\end{equation}
where $B$ is the bank that contains samples from all previous source and target batches, and $b_t$ is the current target batch [13]. By comparing similarities between unlabeled target samples from the current target batch to all elements stored in the bank, current target samples are pushed towards source and target samples with which they share the most similar features. 

\textbf{Entropy Separation (ES) Loss}: The objective of the ES loss is to align classes present in both datasets, while pushing away the classes present only in one of the domains [13]. This is possible because unknown samples often do not share features with known samples, which leads to larger entropies for unknown samples compared to entropies between shared classes [22]. Therefore, the entropy can be used to decide the boundary between known and unknown samples. If we denote the mean entropy of the classifier output $\vectorbold p_i$ of sample $i$ from the target batch $b_t$ as $H(\vectorbold p_i)$, we can define a boundary $\rho$ around the entropy value so that ${\cal L}_\mathrm{ES}(\vectorbold p_i) =
-|H(\vectorbold p_i)-\rho|$ when $|H(\vectorbold p_i)-\rho| > m$, and ${\cal L}_\mathrm{ES}=0$ otherwise.
Here $m$ is a confidence threshold around the boundary $\rho$, which is used to decide if we are confident about whether a particular sample belongs in a known or unknown class. Only those samples that are far enough from the entropy boundary $\rho$ will be moved towards known classes or pushed away as an unknown class. The boundary $\rho$ and confidence threshold $m$ start from preset values (determined from experiments), but are actively fine-tuned during training. Finally, the total ES loss is:
\begin{equation}
{\cal L}_\mathrm{ES} = \frac{1}{|b_t|}\sum_{i \in b_t} {\cal L}_\mathrm{ES}(\vectorbold p_i). 
\end{equation}

\textbf{Total Loss:} The main classification loss for the labeled source domain is the weighted cross-entropy (CE) loss. Combining this with the other terms described above gives the total loss, which is the objective of the model training:
${\cal L} = {\cal L}_{CE} + \lambda({\cal L}_\mathrm{AC} + {\cal L}_\mathrm{ES})$.
The importance of the DA loss terms is governed by the weight parameter $\lambda$. We find that $\lambda =0.005$ achieves the best model performance on our datasets. 

We use \textit{ResNet50} [8] network and train it with early stopping that monitors the change in accuracy and stops the training when there is no improvement in 12 epochs. Domain-specific batch normalization is used to eliminates domain style information leakage. The model is trained using stochastic gradient descent with Nesterov momentum [14] and an initial learning rate of $0.001$ (with an inverse learning rate scheduler, whereby the learning rate is decayed by a factor of $0.1$ every 10 epochs). We train our models on 4 NVIDIA RTX A6000 GPUs (available from Google Colab and LambdaLabs), and on average the training converges in ${\approx}5$ hours. 

\section{Data and Experiments}\label{sec:data}
In this work we apply our method to 3-class and 10-class experiments. We use two datasets from the Galaxy Zoo project [GZ; 12, 19], which used crowd sourcing to create labels for millions of galaxies through a web-based interface: source domain dataset from GZ2 SDSS [19] images, and target domain dataset from GZ3 DECaLS\footnote{Current publicly available GZ datasets can be found at \url{https://data.galaxyzoo.org}} [17], that uses DESI Legacy Imaging Surveys [6]. Specifically, we use a subset of this dataset (with images that passed more rigorous vote filtering) named Galaxy10 DECaLS\footnote{Galaxy10 data is available at \url{https://astronn.readthedocs.io/en/latest/galaxy10.html}}. For our 10-class experiment, we use 9 classes from this dataset (disturbed, merging, round smooth, cigar shaped smooth, barred spiral, unbarred tight spiral, unbarred loose spiral, edge-on without bulge, edge-on with bulge) and add one more class from the full GZ3 DECaLS dataset, gravitationally lensed galaxies, which we will treat as an unknown class. We use the same labels for our source GZ2 SDSS dataset. Furthermore, we also test our method on a simple 3-class problem with spiral, elliptical (known classes), and merging galaxies (unknown class). These correspond, respectively, to the barred spiral, round smooth and merging classes from the 10-class experiment. In Figure 1 we show example images from both source and target domains. In this work, we demonstrate the performance of our model on two \textit{Open DA} problems, where the unlabeled target domain contains one unknown class, not present in the labeled source domain. Both SDSS and DECaLS data include three filter images ($i$,~$r$,~$g$). We use SExtractor [1] to determine the center and radius of objects in the downloaded images, and then crop images to $256\times256$ pixels, using the extracted object properties to ensure no pertinent parts of the image have been inappropriately removed. The source and target datasets each contain ${\approx}6.5$k images for our 3-class problem, and ${\approx}20$k images for out 10-class problem. All data in our experiments are normalized to pixel values $\left[0,1\right]$ and then divided into training, validation, and test sets in proportions $60\%:20\%:20\%$.

\begin{figure}[ht]
    \centering
	\includegraphics[width=0.9\linewidth]{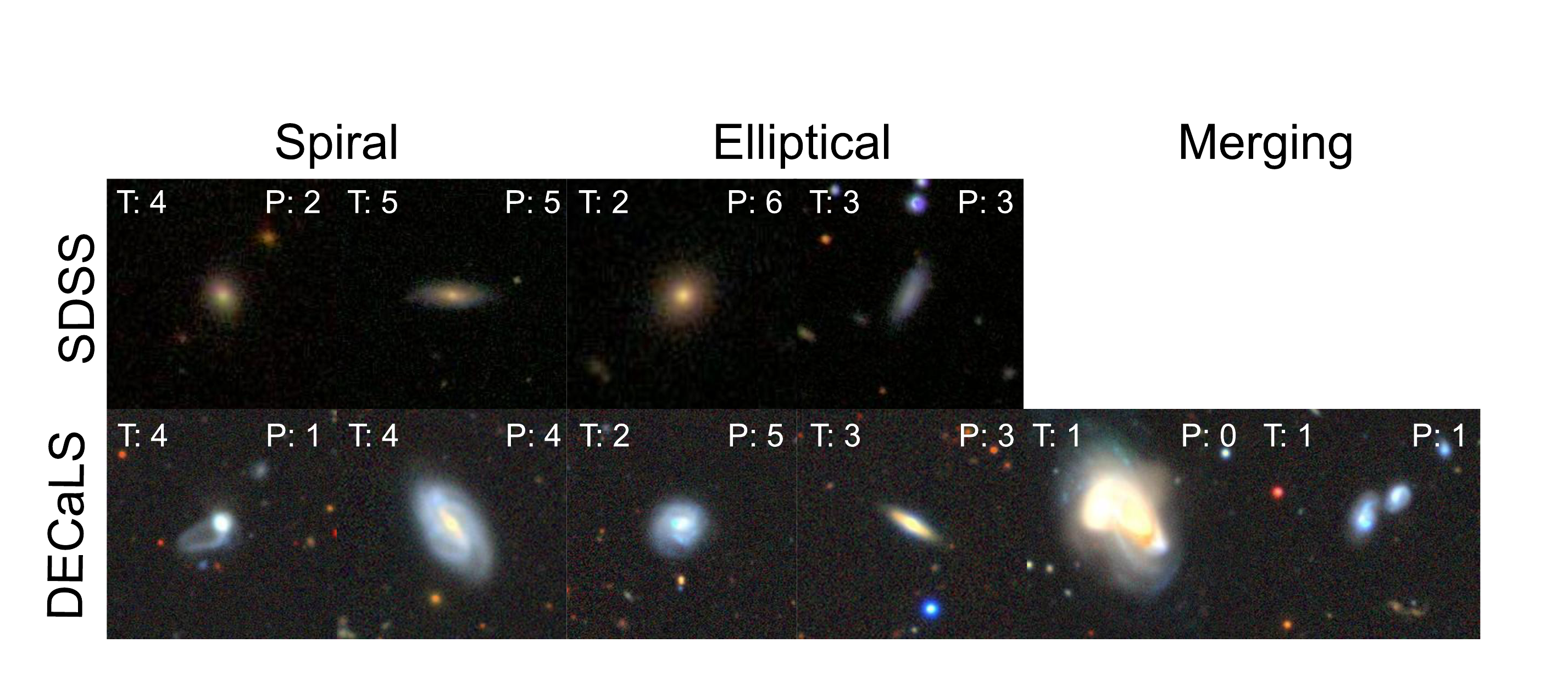}\\
    \caption{Example images from the 3-class problem. Source domain SDSS data (spiral and elliptical classes) is given in the top row, while the target domain DECaLS data (spiral, elliptical, and merging classes) is in the bottom row. We give true labels in the top left corner of each image and predicted labels in the top right corner (when trained with DA). These examples show the most often correctly and incorrectly classified objects in our experiments.}
    \label{fig:examples}
\end{figure}

\section{Results}\label{sec:results}

In Table 1 we report accuracies for the source and target test sets: normal training without any DA, i.e. training with just ${\cal L}_{CE}$ on the source domain (top row), and then \textit{DeepAstroUDA} training with DA (bottom row). Using DA improves performance in both data domains. Aligning the data distributions allows the model to work well on unlabeled target data. The use of domain-invariant features makes training harder, preventing the model from overfitting on the source domain to allow it to train for longer and increase performance even on the labeled source data. In the top row of Figure 2, we show how the target domain accuracies in the 3-class problem change during training, as well as how the three different loss terms evolve. The bottom row shows the training target accuracies for the 10-class problem and a target test set confusion matrix that indicates the similarities between some known galaxy classes. Finally, in Figure 3, we show t-SNE plots [15] of the latent space of the model trained on the 3-class problem, to better illustrate the effects of DA training. Our results show that \textit{DeepAstroUDA} can: 1) successfully be used on difficult domain shift problems such as astronomical data originating from different surveys, and 2) handle any type of domain overlap and perform in the presence of unknown classes, which can be used for anomaly detection tasks like searching for merging galaxies, gravitational lenses, etc. While we focus on cross-survey \textit{Open DA} problems (unknown class present in the target domain) in this paper, we will explore and present results of our method on different types of domain shift problems in our future work.

\begin{table}
   \centering
   \noindent\begin{minipage}[!h]{\linewidth}
   \centering
    \caption{Accuracies for \textit{ResNet50} on SDSS (source) and DECaLS (target) test data (3-class and 10-class problem) when training without DA (top row) and with DA (bottom row). Inclusion of DA increases accuracy for both source and target data in both experiments.
    }
  \label{tab:3_results}
  \centering
  \begin{tabular}{|l |c c| c c|}
 \multicolumn{1}{c}{}     &    \multicolumn{2}{c}{3-class}  &   \multicolumn{2}{c}{10-class} \\
 \hline    Train. type    &  Source  & Target   &  Source  & Target \\\hline
No DA               &  $0.81$       &  $0.56$  &   $0.77$       &  $0.43$   \\ 
DA         &   $\bm{0.84}$       &  $\bm{0.82}$   &   $\bm{0.82}$       &  $\bm{0.79}$  \\ \hline
\end{tabular}
\end{minipage}
\end{table}

\begin{figure}[ht]
    \centering	\includegraphics[width=0.95\linewidth]{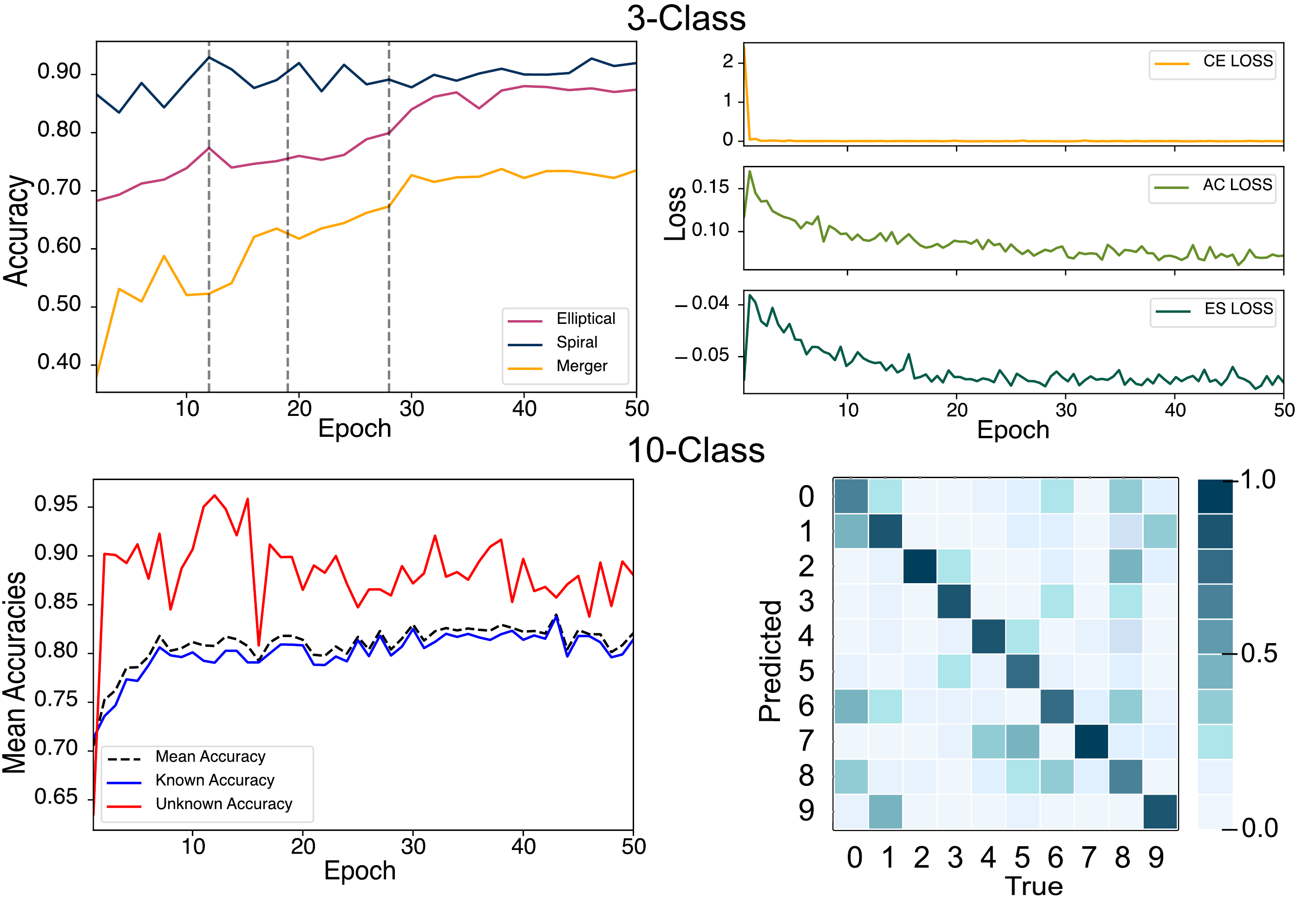}\\
    \caption{\textit{Top (3-class}): left: target domain accuracies during training (elliptical in violet, spiral in navy and unknown merger class in yellow); right: loss functions during training (CE loss in yellow, AC loss in green, ES loss in dark green). Vertical dashed lines on the left plot show epochs in which hyperparameters of the ES loss were fine-tuned. \textit{Bottom (10-class)}: left: target domain mean known class accuracy (blue), unknown gravitational lens class accuracy (red), and mean accuracy for all classes (black dashed line); right: target test set confusion matrix to illustrate confusion between morphologically similar classes (disturbed (0), merging (1), round smooth (2), cigar shaped smooth (3), barred spiral (4), unbarred tight spiral (5), unbarred loose spiral (6), edge-on without bulge (7), edge-on with bulge (8), lenses (9)).}
    \label{fig:3_acc_loss}
\end{figure}
\begin{figure}[!ht]
    \centering
	\includegraphics[width=0.9\linewidth]{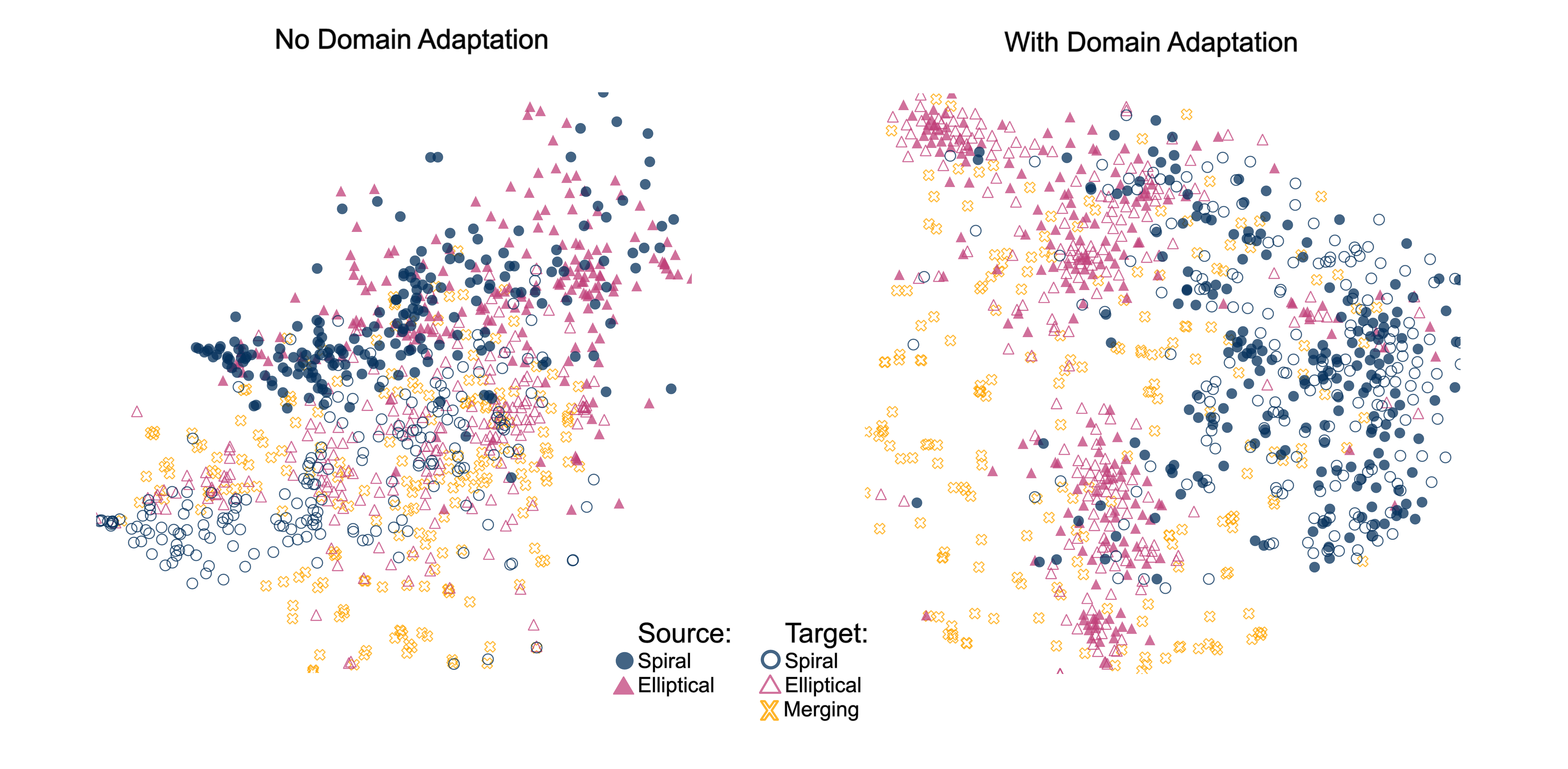}\\
    \caption{t-SNE plots of the latent space of the model trained on the 3-class problem. Without DA (left) the target domain (empty circles) remains completely separate from the source (filled circles), while using DA (right) aligns classes correctly, which allows the model to perform well in both domains. Furthermore, we can see that the unknown merger class (yellow) in the target domain is separated from the rest and moved to the outskirts.}
    \label{fig:tsne}
\end{figure}

\section*{Broader Impact}

Our paper introduces a universal DA method that can handle any type of overlap between different astronomical datasets and is applicable not only to multiple types of problems in astrophysics and cosmology, but also to other scientific domains and beyond. This research will impact the wider scientific community, since work with data from simulations and multiple telescopes or experiments is often present in many scientific applications. This is also relevant for the development of methods trained on simulations or old observations that work during new observations in real time, which is crucial for fast detection of transient phenomena, anomalies or real-time data cataloging and dataset creation.

\begin{ack} 

This manuscript has been supported by Fermi Research Alliance, LLC under Contract No. DE-AC02-07CH11359 with the U.S.\ Department of Energy (DOE), Office of Science, Office of High Energy Physics. This research has been partially supported by the High Velocity Artificial Intelligence grant as part of the DOE High Energy Physics Computational HEP program. 
It has also been partially supported by the DOE Office of Science, Office of Advanced Scientific Computing Research, applied mathematics and SciDAC programs under Contract No.\ DE-AC02-06CH11357. 
This research used resources of the Argonne Leadership Computing Facility at Argonne National Laboratory, which is a user facility supported by the DOE Office of Science.

The authors of this paper have committed themselves to performing this work in an equitable, inclusive, and just environment, and we hold ourselves accountable, believing that the best science is contingent on a good research environment.
We acknowledge the Deep Skies Lab as a community of multi-domain experts and collaborators who have facilitated an environment of open discussion, idea-generation, and collaboration. This community was important for the development of this project.

Furthermore, we also thank the anonymous referees who helped improve this manuscript.

\textbf{Author Contributions:} A.~\'Ciprijanovi\'c: \textit{Conceptualization, Data curation, Formal analysis, Investigation, Methodology, Project administration, Resources, Software, Supervision, Visualization, Writing of original draft}; 
A.~Lewis: \textit{Formal analysis, Investigation, Methodology, Resources, Software, Visualization, Writing of original draft}; 
K.~Pedro: \textit{Conceptualization, Methodology, Project administration, Resources, Software, Supervision, Writing  (review \& editing)}; 
S.~Madireddy: \textit{Conceptualization, Methodology, Resources, Software, Supervision, Writing (review \& editing)}; 
B.~Nord: \textit{Methodology, Supervision, Writing (review \& editing)}; 
G.~N.~Perdue: \textit{Conceptualization, Methodology, Project administration, Resources, Software, Supervision, Writing  (review \& editing)}; 
S.~M.~Wild: \textit{Conceptualization, Methodology, Writing (review \& editing)}.
\end{ack}

\medskip
\small


\medskip
\small
\section*{References}
[1] E. Bertin and S. Arnouts. SExtractor: Software for source extraction. A\&AS, 117:393–404, June 1996.

[2] A.  \'Ciprijanovi \'c, D. Kafkes, K. Downey, S. Jenkins, G. N. Perdue, S. Madireddy, T. Johnston, G. F. Snyder, and B. Nord. DeepMerge - II. Building robust deep learning algorithms for merging galaxy identification
across domains. MNRAS, 506(1):677–691, September 2021.

[3] Aleksandra  \'Ciprijanovi \'c, Diana Kafkes, Gregory Snyder, F. Javier Sánchez, Gabriel Nathan Perdue, Kevin Pedro, Brian Nord, Sandeep Madireddy, and Stefan M. Wild. DeepAdversaries: examining the robustness
of deep learning models for galaxy morphology classification. Machine Learning: Science and Technology, 3(3):035007, September 2022.

[4] Gabriela Csurka. A comprehensive survey on domain adaptation for visual applications. In Domain Adaptation in Computer Vision Applications, pages 1–35. Springer International Publishing, Cham, 2017.

[5] Dark Energy Survey Collaboration, T. Abbott, F. B. Abdalla, J. Aleksi \'c, S. Allam, A. Amara, D. Bacon, E. Balbinot, M. Banerji, and et al. The Dark Energy Survey: more than dark energy - an overview. MNRAS, 460(2):1270–1299, August 2016.

[6] DESI Collaboration. Overview of the DESI Legacy Imaging Surveys. AJ, 157(5):168, May 2019.

[7] Kai Han, Sylvestre-Alvise Rebuffi, Sebastien Ehrhardt, Andrea Vedaldi, and Andrew Zisserman. Automatically Discovering and Learning New Visual Categories with Ranking Statistics. arXiv e-prints, page arXiv:2002.05714, February 2020.

[8] Kaiming He, Xiangyu Zhang, Shaoqing Ren, and Jian Sun. Deep residual learning for image recognition. In 2016 IEEE Conference on Computer Vision and Pattern Recognition (CVPR), pages 770–778, 2016.

[9] \v{Z}eljko Ivezi\'c, Steven M. Kahn, J. Anthony Tyson, Bob Abel, Emily Acosta, Robyn Allsman, David Alonso, Yusra AlSayyad, and et al. LSST: From science drivers to reference design and anticipated data products. ApJ, 873(2):111, March 2019.

[10] Jichang Li, Guanbin Li, Yemin Shi, and Yizhou Yu. Cross-domain adaptive clustering for semi-supervised domain adaptation. In Proceedings of the IEEE/CVF Conference on Computer Vision and Pattern Recognition (CVPR), pages 2505–2514, June 2021.

[11] Chris Lintott, Kevin Schawinski, Steven Bamford, An\v{z}e Slosar, Kate Land, Daniel Thomas, Edd Edmondson, Karen Masters, Robert C. Nichol, M. Jordan Raddick, Alex Szalay, Dan Andreescu, Phil Murray, and Jan Vandenberg. Galaxy Zoo 1: data release of morphological classifications for nearly 900000 galaxies. MNRAS, 410(1):166–178, January 2011.

[12] Chris J. Lintott, Kevin Schawinski, An\v{z}e Slosar, Kate Land, Steven Bamford, Daniel Thomas, M. Jordan Raddick, Robert C. Nichol, Alex Szalay, Dan Andreescu, Phil Murray, and Jan Vandenberg. Galaxy Zoo: morphologies derived from visual inspection of galaxies from the Sloan Digital Sky Survey. MNRAS, 389(3):1179–1189, September 2008.

[13] Kuniaki Saito, Donghyun Kim, Stan Sclaroff, and Kate Saenko. Universal Domain Adaptation through Self Supervision. arXiv e-prints, page arXiv:2002.07953, February 2020.

[14] Ilya Sutskever, James Martens, George Dahl, and Geoffrey Hinton. On the importance of initialization and momentum in deep learning. In Sanjoy Dasgupta and David McAllester, editors, Proceedings of the 30th International Conference on Machine Learning, volume 28 of Proceedings of Machine Learning Research, pages 1139–1147, Atlanta, Georgia, USA, 17–19 Jun 2013. PMLR.

[15] L. van der Maaten and G. Hinton. Visualizing data using t-SNE. Journal of Machine Learning Research, 9(86):2579–2605, 2008.

[16] Ricardo Vilalta, Kinjal Dhar Gupta, Dainis Boumber, and Mikhail M. Meskhi. A General Approach to Domain Adaptation with Applications in Astronomy. PASP, 131(1004):108008, October 2019.

[17] Mike Walmsley, Chris Lintott, Tobias G\'eron, Sandor Kruk, Coleman Krawczyk, Kyle W. Willett, Steven Bamford, Lee S. Kelvin, Lucy Fortson, Yarin Gal, William Keel, Karen L. Masters, Vihang Mehta, Brooke D. Simmons, Rebecca Smethurst, Lewis Smith, Elisabeth M. Baeten, and Christine Macmillan. Galaxy Zoo DECaLS: Detailed visual morphology measurements from volunteers and deep learning for 314 000 galaxies. MNRAS, 509(3):3966–3988, January 2022.

[18] Mei Wang and Weihong Deng. Deep visual domain adaptation: A survey. Neurocomputing, 312:135–153, 2018.

[19] Kyle W. Willett, Chris J. Lintott, Steven P. Bamford, Karen L. Masters, Brooke D. Simmons, Kevin R. V. Casteels, Edward M. Edmondson, Lucy F. Fortson, Sugata Kaviraj, William C. Keel, Thomas Melvin, Robert C. Nichol, M. Jordan Raddick, Kevin Schawinski, Robert J. Simpson, Ramin A. Skibba, Arfon M. Smith, and Daniel Thomas. Galaxy Zoo 2: detailed morphological classifications for 304 122 galaxies from the Sloan Digital Sky Survey. MNRAS, 435(4):2835–2860, November 2013.

[20] Garrett Wilson and Diane J. Cook. A survey of unsupervised deep domain adaptation. ACM Transactions on Intelligent Systems and Technology, 11(5), July 2020.

[21] Donald G. York, J. Adelman, Jr. Anderson, John E., Scott F. Anderson, James Annis, Neta A. Bahcall, J. A. Bakken, Robert Barkhouser, Steven Bastian, Eileen Berman, et al., and SDSS Collaboration. The Sloan
Digital Sky Survey: Technical Summary. AJ, 120(3):1579–1587, September 2000.

[22] Kaichao You, Mingsheng Long, Zhangjie Cao, Jianmin Wang, and Michael I. Jordan. Universal domain adaptation. In 2019 IEEE/CVF Conference on Computer Vision and Pattern Recognition (CVPR), pages 2715–2724, 2019.

\section*{Checklist}
\begin{enumerate}

\item For all authors...
\begin{enumerate}
  \item Do the main claims made in the abstract and introduction accurately reflect the paper's contributions and scope?
    \answerYes{}
  \item Did you describe the limitations of your work?
    \answerNo{We do not discuss limitations in this short preliminary study, but we will discuss it in our more detailed future work.}
  \item Did you discuss any potential negative societal impacts of your work?
    \answerNA{We do not see any obvious negative social impact of our work, which is related to domain adaptation for astronomy.}
  \item Have you read the ethics review guidelines and ensured that your paper conforms to them?
    \answerYes{}
\end{enumerate}

\item If you are including theoretical results...
\begin{enumerate}
  \item Did you state the full set of assumptions of all theoretical results?
    \answerNA{We do not include any theoretical results.}
        \item Did you include complete proofs of all theoretical results?
    \answerNA{}
\end{enumerate}

\item If you ran experiments...
\begin{enumerate}
  \item Did you include the code, data, and instructions needed to reproduce the main experimental results (either in the supplemental material or as a URL)?
    \answerYes{We provide URL for our GitHub repository, where the code used in this work is publicly available.}
  \item Did you specify all the training details (e.g., data splits, hyperparameters, how they were chosen)?
    \answerYes{Details are given in Section 2 and Section 3.}
        \item Did you report error bars (e.g., with respect to the random seed after running experiments multiple times)?
    \answerNo{In this preliminary study we only report results from a single run without multiple random seeds.}
        \item Did you include the total amount of compute and the type of resources used (e.g., type of GPUs, internal cluster, or cloud provider)?
    \answerYes{We give details in Section 2.}
\end{enumerate}

\item If you are using existing assets (e.g., code, data, models) or curating/releasing new assets...
\begin{enumerate}
  \item If your work uses existing assets, did you cite the creators?
    \answerYes{In Section 3 we reference papers related to the datasets we use, as well as provide links where the data is available.}
  \item Did you mention the license of the assets?
    \answerNA{}
  \item Did you include any new assets either in the supplemental material or as a URL?
    \answerNo{}
  \item Did you discuss whether and how consent was obtained from people whose data you're using/curating?
    \answerNA{We are using publicly available Galaxy Zoo data.}
  \item Did you discuss whether the data you are using/curating contains personally identifiable information or offensive content?
    \answerNA{Astronomical data does not contain any personally identifiable information.}
\end{enumerate}

\item If you used crowdsourcing or conducted research with human subjects...
\begin{enumerate}
  \item Did you include the full text of instructions given to participants and screenshots, if applicable?
    \answerNA{We did not use any crowdsourcing. Even though datasets were labeled via crowdsourcing, all work was performed beforehand, through the Galaxy Zoo project, which we were not involved in.}
  \item Did you describe any potential participant risks, with links to Institutional Review Board (IRB) approvals, if applicable?
    \answerNA{}
  \item Did you include the estimated hourly wage paid to participants and the total amount spent on participant compensation?
    \answerNA{}
\end{enumerate}

\end{enumerate}






\end{document}